\documentclass[conference]{IEEEtran}
\IEEEoverridecommandlockouts
\usepackage{amsmath,amssymb,amsfonts}
\usepackage{algorithmic}
\usepackage{graphicx}
\usepackage{textcomp}
\usepackage{url}
\usepackage{xcolor}
\usepackage{array}
\usepackage{adjustbox}
\usepackage{graphicx}
\usepackage{adjustbox}
\usepackage{afterpage}
\usepackage{pdfpages}

\newenvironment{purple}{
    \color{purple}
}{
    \ignorespacesafterend
}

\begin{document}

\title{Leveraging Retrieval-Augmented Generation for Persian University Knowledge Retrieval\\

\thanks{We extend our gratitude to our fellow group members within the UIAI Community at the University of Isfahan—Kiana Fakhrian, Amirhossein Moalemi, Amirhossein Ala, and Zahra Mortazavi—for their dedicated efforts in integrating our dataset and diligently scraping documents from the University of Isfahan website. Their contributions were instrumental to the progress of this work.}
}

\author{\IEEEauthorblockN{1\textsuperscript{st} Arshia Hemmat}
\IEEEauthorblockA{\textit{dept. Computer Engineering} \\
\textit{University of Isfahan}\\
Isfahan, Iran \\
arshiahemmat@mehr.ui.ac.ir}
\and
\IEEEauthorblockN{1\textsuperscript{st} Kianoosh Vadaei}
\IEEEauthorblockA{\textit{dept. Computer Engineering} \\
\textit{University of Isfahan}\\
Isfahan, Iran \\
k.vadaei@mehr.ui.ac.ir}
\and
\IEEEauthorblockN{1\textsuperscript{st} Mohammad Hassan Heydari}
\IEEEauthorblockA{\textit{dept. Computer Engineering} \\
\textit{University of Isfahan}\\
Isfahan, Iran \\
mheydarii@mehr.ui.ac.ir}

\and
\IEEEauthorblockN{2\textsuperscript{nd} Afsaneh Fatemi}
\IEEEauthorblockA{\textit{dept. Computer Engineering} \\
\textit{University of Isfahan}\\
Isfahan, Iran \\
a\_fatemi@eng.ui.ac.ir}
}

\maketitle

\begin{abstract}
\label{sec:abstract}
This paper introduces an innovative approach using Retrieval-Augmented Generation (RAG) pipelines with Large Language Models (LLMs) to enhance information retrieval and query response systems for university-related question answering. By systematically extracting data from the university's official website, primarily in Persian, and employing advanced prompt engineering techniques, we generate accurate and contextually relevant responses to user queries.

We developed a comprehensive university benchmark, UniversityQuestionBench (UQB), to rigorously evaluate our system’s performance. UQB focuses on Persian-language data, assessing accuracy and reliability through various metrics and real-world scenarios. Our experimental results demonstrate significant improvements in the precision and relevance of generated responses, enhancing user experiences, and reducing the time required to obtain relevant answers. 

In summary, this paper presents a novel application of RAG pipelines and LLMs for Persian-language data retrieval, supported by a meticulously prepared university benchmark, offering valuable insights into advanced AI techniques for academic data retrieval and setting the stage for future research in this domain.\footnote{Dataset is publicly available at \url{https://huggingface.co/datasets/UIAIC/UQB}}

\end{abstract}


\section{Introduction}
\label{sec:intro}
Large Language Models (LLMs), including cutting-edge ones like OpenAI GPTs and Google Gemini models, often face significant challenges when it comes to extracting and utilizing local data, particularly from specialized datasets such as universities archives. These models are typically trained on broad, diverse datasets, which can result in a lack of specificity and accuracy when applied to niche domains. The challenges include the inability to access and process localized data effectively, leading to issues like hallucinations and inaccuracies in generated content. Additionally, the models' reliance on pre-existing knowledge limits their capability to incorporate newly acquired, domain-specific information without extensive retraining \cite{lewis2021retrievalaugmentedgenerationknowledgeintensivenlp, zhao2024}.

\begin{figure*}
  \clearpage
  \centering
  \includegraphics[width=0.96\textwidth]{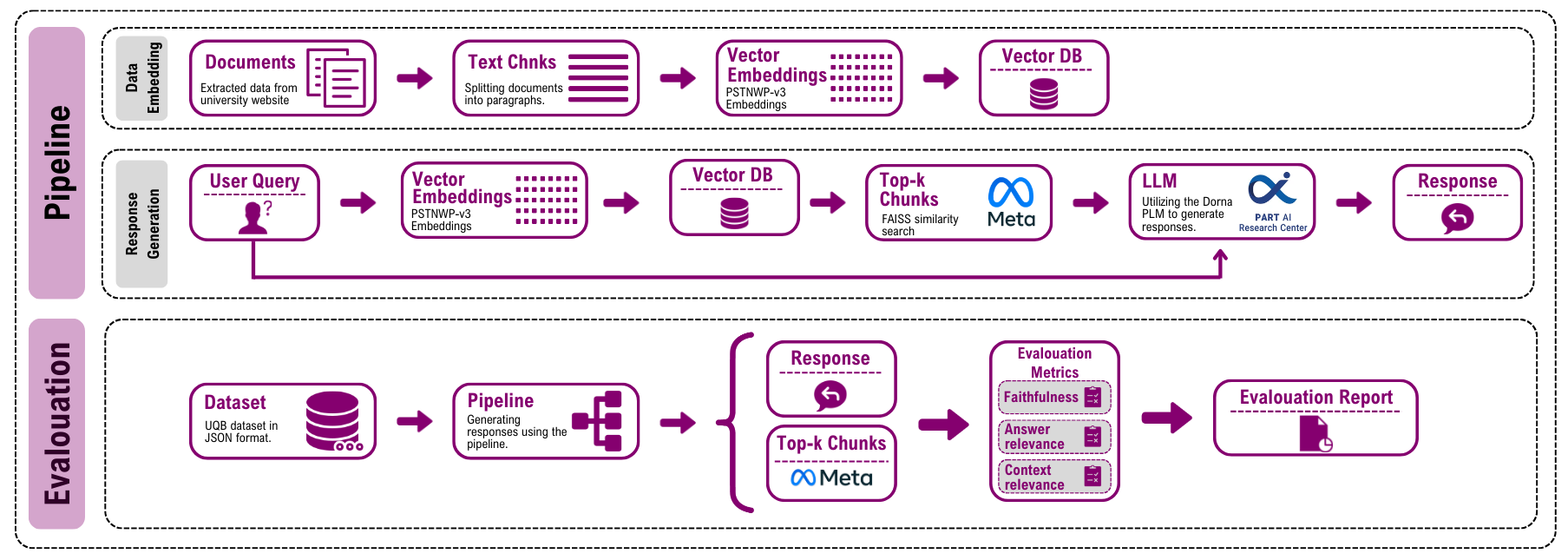} 
  \caption{Our Proposed Pipeline }
  \label{fig:demo_plot}
\end{figure*}

Retrieval-Augmented Generation (RAG) offers a robust solution to the challenges faced by LLMs in processing local documents. By integrating retrieval mechanisms with generation capabilities, RAG pipelines enable models to access and utilize specific, relevant information from extensive datasets. Our proposed pipeline leverages a two-stage RAG approach combined with a Persian Large Language Model (PLM) and advanced prompt engineering techniques. Initially, queries are categorized to identify the most relevant documents, after which the appropriate LLM is engaged to generate accurate and contextually relevant responses. This method significantly enhances the precision and utility of LLMs in handling localized, domain-specific queries \cite{NeurIPS2020, siriwardhana2022improvingdomainadaptationretrieval}.

We developed the "UniversityQuestionBench" dataset, created from the most frequently asked questions by students across various disciplines. This dataset is designed to evaluate the performance of Persian LLMs integrated with RAG using the RAGAS evaluation metrics, which includes three key measures: Faithfulness, Answer Relevance and Context Relevance. By employing these metrics, we ensure that the model provides accurate, relevant, and contextually appropriate responses. The dataset and evaluation processes aim to benchmark the effectiveness of our pipeline in addressing the specific needs of universities students \cite{es2023ragasautomatedevaluationretrieval, yu-2022-retrieval, chen2023benchmarkinglargelanguagemodels}.

Our contributions to this paper are as follows:
\begin{itemize}
  \item Development of a two-stage RAG pipeline integrated with Persian LLMs for handling localized queries.
  \item Creation of the UniversityQuestionBench dataset, tailored to the most common queries from university students.
  \item Leveraging the RAGAS evaluation metrics to rigorously assess the performance of our models.
  \item Demonstration of significant improvements in Faithfulness, Answer Relevance and Context Relevance of Responses generated by our pipeline.
\end{itemize}

\section{Related Work}
\label{sec:related}
\subsection{Introduction to Retrieval-Augmented Generation}

Retrieval-Augmented Generation (RAG) is a novel paradigm that enhances the performance of large language models by incorporating information retrieval processes into the generation mechanism. This approach aims to improve the accuracy and robustness of generated content by utilizing relevant external data sources. Recent studies have demonstrated the effectiveness of RAG frameworks in various applications in AI and machine learning ~\cite{lewis2021retrievalaugmentedgenerationknowledgeintensivenlp, gao2024retrievalaugmentedgenerationlargelanguage}.

\subsection{Recent Advances and Techniques in RAG}

Recent advancements in RAG have focused on innovative techniques and methodologies to optimize retrieval and generation processes. Lewis et al. (2020) highlight the power of RAG in knowledge-intensive NLP tasks, demonstrating its potential to solve complex information retrieval challenges \cite{lewis2021retrievalaugmentedgenerationknowledgeintensivenlp}. Shahul et al. (2023) introduced RAGAS, a framework for automated evaluation of RAG pipelines, emphasizing the importance of reference-free evaluation metrics to enhance the evaluation process of RAG systems. Siriwardhana et al. (2022) developed RAG-end2end, which optimizes RAG for domain-specific knowledge bases, significantly improving performance in specialized domains such as healthcare and news \cite{siriwardhana2022improvingdomainadaptationretrieval}. Yu (2022) explored the use of retrieval-augmented generation across heterogeneous knowledge, addressing the challenges of retrieving information from diverse sources \cite{yu-2022-retrieval}. Nakhod (2023) proposed applying RAG to elevate low-code developer skills by integrating domain-specific knowledge into large language models, thereby improving their practical utility \cite{nakhod2023using}. Melz (2023) introduced ARM-RAG, a system that enhances large language models' intelligence through storing and retrieving reasoning chains, demonstrating significant improvements in problem-solving tasks \cite{melz2023enhancingllmintelligencearmrag}. Chen et al. (2023) provided a comprehensive evaluation of the impact of RAG on large language models, highlighting the potential bottlenecks and challenges in applying RAG across different tasks \cite{chen2023benchmarkinglargelanguagemodels}. Heydari et al. (2024) proposed the Context Awareness Gate (CAG) architecture, a novel mechanism that dynamically adjusts the LLM’s input prompt based on whether the user query necessitates external context retrieval, thereby enhancing the efficiency and accuracy of RAG systems \cite{heydari2024context}.

\subsection{Applications and Case Studies}

The versatility of RAG is evident in its wide range of applications. For example, RAG has been successfully applied to AI-generated content, enhancing the quality and contextual relevance of the outputs. Another significant application is the integration of RAG across heterogeneous knowledge bases, which has proven effective in generating coherent and contextually appropriate responses \cite{ACLAnthology2022}. Practical applications in high-performance computing for code development further demonstrate its versatility \cite{NVIDIA2024}. Specifically, Graph-based approaches like GNN-RAG and KG-RAG have significantly improved handling complex queries and enhancing factual consistency \cite{Mavromatis2024, Sanmartin2024}.

\subsection{Frameworks and Implementations}

Several frameworks and implementations have been proposed to facilitate the deployment of RAG systems. The University of Massachusetts introduced Stochastic RAG, an end-to-end framework that leverages stochastic methods for retrieval and generation, ensuring high relevance and diversity in the outputs \cite{UMass2022}. Additionally, the Spring AI project demonstrates the practical application of RAG using Azure OpenAI, providing valuable insights into the integration of RAG with cloud-based AI services \cite{SpringAI2022}. The Semantic Kernel framework by Microsoft offers another robust implementation for RAG \cite{DotNet2023}. KRAGEN, a knowledge graph-enhanced RAG framework, has also been developed for biomedical problem-solving, illustrating the application of RAG in specialized domains \cite{Matsumoto2024}.

\subsection{Leveraging RAG in Specific Domain Tasks}

RAG has shown significant potential in addressing specific domain tasks, such as enhancing university information systems. By crawling university websites and creating datasets tailored to students' queries, RAG can effectively answer questions related to different departments and services. This approach not only improves the accuracy of information retrieval but also enhances the overall user experience by providing precise and relevant answers to specific queries \cite{NVIDIA2024, AtomCamp2024}. Graph-based approaches have further enhanced RAG's capabilities in handling domain-specific tasks by integrating structured knowledge representations \cite{Edge2024, Dong2024}.

\subsection{Challenges and Future Directions}

Despite the promising advancements, several challenges remain in the implementation of RAG. A survey by Semantic Scholar identifies key obstacles, including the complexity of integrating retrieval mechanisms with generation models and ensuring the scalability of such systems \cite{SemanticScholar2024}. The tutorial by IJCAI further discusses recent advances and outlines future research directions to overcome these challenges, emphasizing the need for continued innovation in this field \cite{IJCAI2022}.

\section{Dataset Generation}
\label{sec:dataset_generation}

\subsection{Data Production Process}

The data generation process for our UniversityQuestionBench dataset involves several key steps to ensure the collection of relevant and frequently asked questions by students. We initiated the process by identifying the most important data from the University of Isfahan's website.\footnote{University of Isfahan website: \url{https://ui.ac.ir}}
. This data extraction step is crucial for forming the foundational dataset.
%Here we should reference the Image of dataset production. 

\vspace{1em}

\textbf{\textit{Step 1:}}
We start by collecting data $\mathcal{D}$ from the university's official website, focusing on the most critical and frequently accessed information by students. The first step is represented in the Equation \ref{eq:data}:
\begin{equation}
\label{eq:data}
\mathcal{D} = \{\text{data}_{i}\}_{i=1}^N
\end{equation}

where $\text{data}_{i}$ represents individual pieces of information and $N$ is the total number of data items extracted.

Next, we surveyed students to gather insights on the most common questions they encounter. The survey results were recorded and analyzed to identify patterns and frequently asked questions.

\vspace{1em}

\textbf{\textit{Step 2:}}
In the second stage, we collected the questions from students through two primary approaches to ensure comprehensive coverage of frequently asked questions:

\begin{enumerate}
    \item \textbf{Student Surveys:}We utilized data collected from 60 students to refine and enhance the quality of the documents, ensuring they accurately reflected the most relevant and practical information based on real student experiences and inquiries. We asked students to fill out a form regarding the most frequent questions they encounter at the university. The form included open-ended sections where students could describe the challenges they face when navigating university resources. From the responses, we identified a correspondence corpus of frequently asked questions that were representative of common student concerns from either the university website or the relevant documents in the different channels.
    
    \item \textbf{Web Scraping:} In addition to the student surveys, we scraped the official Isfahan University website for relevant information. This included extracting questions derived from useful content such as department-specific pages, contact information (e.g., email addresses of professors), course descriptions, and administrative procedures. These questions were designed to complement the student-provided data and fill in any gaps in coverage.
\end{enumerate}

The collected data from these two sources forms the initial set of questions, denoted as $\mathcal{Q}_{\text{student}}$. This step is formulated in Equation \ref{eq:student}:
\begin{equation}
\label{eq:student}
\mathcal{Q}_{\text{student}} = \{\text{question}_{j}\}_{j=1}^M
\end{equation}
where $\text{question}_{j}$ represents each individual question and $M$ is the total number of questions collected.

where $\text{question}_{j}$ represents each individual question and $M$ is the total number of questions collected.

To supplement the student-provided questions, we used GPT-4 to generate additional questions. We implemented a Python script to automate the generation of a comprehensive question set, ensuring a diverse and complete dataset.
\vspace{1em}

\textbf{\textit{Step 3:}}
Let $\mathcal{Q}_{\text{gpt}}$ represent the set of questions generated by GPT-4. This is formulated in the Equation \ref{eq:gpt}:
\begin{equation}
\label{eq:gpt}
\mathcal{Q}_{\text{gpt}} = \{\text{question}_{k}\}_{k=1}^P,
\end{equation}
where $\text{question}_{k}$ represents each individual question generated by GPT-4 and $P$ is the total number of generated questions.
\vspace{1em}
 
\textbf{\textit{Step 4:}}
The combined set of questions, $\mathcal{Q}_{\text{combined}}$, includes both student-provided and GPT-generated questions, this equation can be shown in the Equation \ref{eq:combined}.
\begin{equation}
\label{eq:combined}
\mathcal{Q}_{\text{combined}} = \mathcal{Q}_{\text{student}} \cup \mathcal{Q}_{\text{gpt}},
\end{equation}

To ensure the accuracy and relevance of the answers, we manually curated the responses with the help of human feedback. This iterative process involved verifying and updating the answers based on student feedback.
\vspace{1em}

\textbf{\textit{Step 5:}}
The set of validated question-answer pairs, $\mathcal{QA}_{\text{valid}}$, is established in the Equation \ref{eq:qa}.
\begin{equation}
\label{eq:qa}
\mathcal{QA}_{\text{valid}} = \{(\text{question}_{j}, \text{answer}_{j})\}_{j=1}^{T},
\end{equation}
where $\mathcal{T}$ corresponds to the Total number of GPT and Human Questions.Additionally, $\text{answer}_{j}$ corresponds to the manually validated answer for $\text{question}_{j}$.

\begin{figure}[t]
  \centering
  \includegraphics[width=0.98\linewidth]{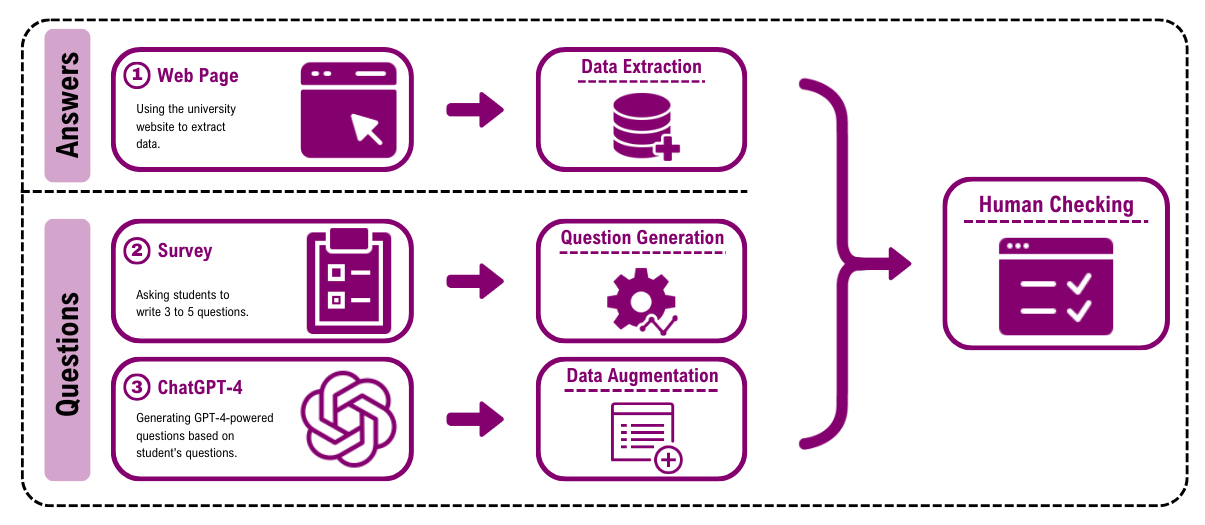} 
  \caption{Data Generation Procedure - In this figure we has shown the question and answer generation.}
  
  \label{fig:datageneration}
\end{figure}

The final dataset comprises over 500 documents, encompassing a wide range of contexts to ensure comprehensive coverage. These documents include detailed information about the academic groups and sections of each department, professors along with their contact emails, and various aspects of the university, such as administrative procedures, facilities, and other relevant resources. This extensive collection ensures that the dataset accurately represents the diverse informational needs of students across the university. We illustrate the process of the Dataset Generation in the figure \ref{fig:datageneration}. 

\section{Pipeline Generation}
\label{sec:approach}

\subsection{Pipeline Setting}

As previously discussed, large language models (LLMs) encounter challenges when responding to queries that they were not well-trained on, particularly due to the absence of specific training data related to those queries. Additionally, some queries directly pertain to local or private datasets. The retrieval-augmented generation (RAG) pipeline offers a solution to this issue without requiring extensive fine-tuning of the entire model on the specific dataset . 
\vspace{2em}

Our pipeline is constructed based on the following steps:

\textbf{\textit{Step 1:}}
Pass the crawled data, $\mathcal{D}$, through the model, regardless of whether the data exists in the question or not. This is represented in the Equation \ref{eq:data_pipe}:
\begin{equation}
\label{eq:data_pipe}
\mathcal{D} = \{\text{data}_{i}\}_{i=1}^N,
\end{equation}
where $\text{data}_{i}$ represents individual pieces of information and $N$ is the total number of data items extracted.

\vspace{1em}

\textbf{\textit{Step 2:}}

Identify the type of question and determine the relevant department using the DORNA Model, which is a fine-tuned version of Llama-3 on Presian data \cite{llama-3, llama-1}. Specifically, we employed the 8-bit quantized version of Dorna utilizing QLoRA \cite{qlora}. This approach enables us to load our base model with significantly reduced memory requirements.
Let $\mathcal{T}_{\text{dept}}(q)$ be the function that assigns a question $q$ to a specific department. This can be represented in the Equation \ref{eq:eq2}:
\begin{equation}
\label{eq:eq2}
\mathcal{T}_{\text{dept}}(q) = \text{DORNA}_{\text{classify}}(q),
\end{equation}

\vspace{1em}

\textbf{\textit{Step 3:}}

Split paragraphs from the texts of each department. Let $\mathcal{P}$ denote the set of paragraphs split from $\mathcal{D}$, The formula can be written in the Equation \ref{eq:eq3}:
\begin{equation}
\label{eq:eq3}
\mathcal{P} = \{\text{para}_{j}\}_{j=1}^M,
\end{equation}
where $\text{para}_{j}$ represents each paragraph and $M$ is the total number of paragraphs.

\vspace{1em}

\textbf{\textit{Step 4:}}

Use FAISS to find the similarity function. FAISS is a library for efficient similarity search and clustering of dense vectors, crucial for retrieving similar texts \cite{faiss_paper}. We utilize the Persian embedding model named persian-sentence-transformer-news-wiki-pairs-v3 for embedding the paragraphs.

Let $\mathcal{E}(\cdot)$ be the embedding function provided by the Persian sentence transformer. The embeddings for the paragraphs are in the Equation \ref{eq4}:
\begin{equation}
\label{eq4}
\mathcal{E}(\mathcal{P}) = \{\mathcal{E}(\text{para}_{j})\}_{j=1}^M,
\end{equation}

\vspace{1em}

\textbf{\textit{Step 5:}}

For a given query $q$, its embedding is denoted as $\mathcal{E}(q)$. We then retrieve the first 3 closest documents based on the text similarity between the question and the retrieved contents. This similarity search is represented in the Equation \ref{eq5}:
\begin{equation}
\label{eq5}
\mathcal{R}(q) = \text{TopK}(\text{FAISS}(\mathcal{E}(q), \mathcal{E}(\mathcal{P})), 3),
\end{equation}
where $\mathcal{R}(q)$ represents the set of top 3 retrieved paragraphs similar to the query $q$.

\vspace{1em}

\textbf{\textit{Step 6:}}

Create a prompt template and pass it to the LLMs. Our LLM, DORNA, is a fine-tuned version on Llama-3 of persian data. The prompt template $\mathcal{T}$ is designed to incorporate the retrieved paragraphs and the query.

\vspace{1em}

\textbf{\textit{Step 7:}}

Pass the generated prompt to the LLMs to produce the final answer. Let $\mathcal{A}$ denote the answer generated by DORNA in the Equation \ref{eq:eq7}:
\begin{equation}
\label{eq:eq7}
\mathcal{A} = \text{DORNA}(\mathcal{T}(q, \mathcal{R}(q))),
\end{equation}

\vspace{1em}

These steps constitute our pipeline, leveraging RAG and advanced embedding techniques to ensure accurate and relevant responses from localized data sources. We illustrate our Pipeline schema in figure \ref{fig:demo_plot}.

\begin{table*}[htbp]
\centering
\caption{Evaluation Metrics for Different Models and Embeddings}
\label{table:results}
\begin{tabular}{|l|l|c|c|c|}
\hline
\textbf{Model} & \textbf{Embedding} & \textbf{Faithfulness} & \textbf{Answer Relevancy} & \textbf{Context Relevancy} \\ \hline
GPT 4o & OpenAI Embeddings& 0.6333 & 0.6395 & 0.1154 \\ \hline
GPT 3.5-turbo & OpenAI Embeddings& \textbf{0.8497} & 0.5604 & 0.1849 \\ \hline
GPT 3.5-turbo & Persin-Sentence-Embedding-V3 & 0.8113 & 0.493 & \textbf{0.223} \\ \hline
GPT 4o & Persin-Sentence-Embedding-V3 & 0.6578 & 0.6564 & 0.1848 \\ \hline
Dorna (Persian version of Llama3) & Dorna Embeddings& 0.839 & \textbf{0.823} & 0.216 \\ \hline
\end{tabular}
\end{table*}

\section{Experiments}
\label{sec:experiments}

To comprehensively assess the effectiveness of our RAG pipeline and LLMs, we utilize three key metrics as defined in the RAGAS paper: Faithfulness, Answer Relevance and Context Relevance. Each metric is described in detail below.

\subsection{Faithfulness}

Faithfulness evaluates how accurately the generated answer reflects the content of the retrieved documents. This metric is crucial to ensure that the model does not introduce hallucinations or incorrect information. Let $\mathcal{F}$ denote pipeline faithfulness, calculated in the equation \ref{eq10}:

\begin{equation}
\label{eq10}
\mathcal{F} = \frac{1}{N} \sum_{i=1}^N \text{Faithfulness}(\mathcal{A}_i, \mathcal{R}(q_i)),
\end{equation}

Where:
\begin{itemize}
    \item $\mathcal{A}_i$ is the answer generated for query $q_i$.
    \item $\mathcal{R}(q_i)$ is the set of documents retrieved for query $q_i$.
    \item $\text{Faithfulness}(\mathcal{A}_i, \mathcal{R}(q_i))$ measures the relevancy of $\mathcal{A}_i$ against the information in $\mathcal{R}(q_i)$.
\end{itemize}

The faithfulness function can be further defined in the equation \ref{eq17}:

\begin{equation}
\label{eq17}
\text{Faithfulness}(\mathcal{A}_i, \mathcal{R}(q_i)) = \frac{\sum_{j=1}^{|\mathcal{A}(i)|} \text{Rel}(\mathcal{A}_{ij} , \mathcal{R}(q_i))}{|\mathcal{A}(i)|}
\end{equation}

where $\mathcal{A}_i$ is the set of statements in the answer and $\text{Rel}(\mathcal{A}_{ij}, \mathcal{R}(q_i))$ determines that the statement $\mathcal{A}_{ij}$ from answer $\mathcal{A}_i$ is supported by the document $\mathcal{R}(q_i)$ or not.

\vspace{1em}

\subsection{Answer Relevance}

Answer relevance measures how well the generated answer addresses the query. This metric ensures that the answer is directly relevant to the question asked. We first initialize a set of questions $q$ that can directly address from the Answer $\mathcal{A}_i$. Let $\mathcal{R}_{\text{ans}}$ denote answer relevance, calculated in the equation \ref{eq2}:

\begin{equation}
\label{eq2}
\mathcal{R}_{\text{ans}} = \frac{1}{m} \sum_{j=1}^m \text{Sim}(\mathcal{Q}, q_j)
\end{equation}

Where:
\begin{itemize}
    \item $m$ is the number of questions in the set of questions $q$
    \item $\mathcal{Q}$ is the users query which model generated $\mathcal{A}_i$ based on it.
    \item $\mathcal{A}_i$ is the answer generated for users query $\mathcal{Q}$.
    \item $\text{Sim}(\mathcal{Q}, q_i)$ calculates the similarity of $\mathcal{Q}$ and $q_j$ embedding vectors. In our study, we used cosine similarity to calculate the output of this function.
\end{itemize}

\vspace{1em}

\subsection{Context Relevance}

Context relevance assesses how well the retrieved documents match the query, ensuring that the documents are contextually appropriate and useful for generating an accurate answer. Let $\mathcal{R}_{\text{ctx}}$ denote context relevance, calculated in the equation \ref{eq3}:

\begin{equation}
\label{eq3}
\mathcal{R}_{\text{ctx}} = \frac{1}{N} \sum_{i=1}^N \text{Relevance}(\mathcal{R}(q_i), q_i)
\end{equation}

Where:
\begin{itemize}
    \item $\mathcal{R}(q_i)$ is the set of documents retrieved for query $q_i$.
    \item $\text{Relevance}(\mathcal{R}(q_i), q_i)$ evaluates how well the set of retrieved documents $\mathcal{R}(q_i)$ addresses the query $q_i$.
\end{itemize}

The relevance function for context is further defined in the equation \ref{faith}:

\begin{equation}
\label{faith}
\text{Relevance}(\mathcal{R}(q_i), q_i) = \frac{\sum_{j=1}^{|\mathcal{R}(q_i)|} \text{P}(\mathcal{R}_j(q_i), q_i)}{|\mathcal{R}(q_i)|}
\end{equation}

where $\text{P}(\mathcal{R}_j(q_i), q_i)$ is a measure which indicates the potential of 
 $\mathcal{R}_j(q_i)$ in the set of retrieved documents $\mathcal{R}(q_i)$ for answering the query $q_i$ ,

\section{Results}
\label{sec:Results}

\subsection{Pipeline Performance On UQB}
Our pipeline’s performance was evaluated on a dataset of 300 questions and answers using the test set of the UniversityQuestionBench (UQB) dataset. We used our base model, Dorna, to compute the evaluation metrics. As previously discussed, the evaluation focused on three key metrics: faithfulness, answer relevance, and context relevance. The calculated results are demonstrated in Table \ref{table:results}.

\begin{itemize}
    \item \textbf{Faithfulness:}
    \begin{itemize}
        \item Faithfulness measures the factual accuracy of the responses, reflecting the system's ability to generate outputs consistent with the underlying data source.
        \item The highest \textbf{faithfulness score (0.8497)} is achieved by \textbf{GPT-3.5-turbo with OpenAI Embeddings}, underscoring the robustness of general-purpose embeddings in generating accurate responses.
        \item The performance of \textbf{Dorna with Dorna Embeddings (0.839)} is competitive, highlighting the capability of localized embeddings specifically designed for Persian-language content.
        \item Models utilizing \textbf{Persian-Sentence-Embedding-V3} (e.g., \textbf{GPT-3.5-turbo, 0.8113}) exhibit slightly reduced faithfulness compared to OpenAI Embeddings, suggesting limitations in the current iteration of these embeddings for ensuring strict fidelity to source information.
    \end{itemize}

    \item \textbf{Answer Relevancy:}
    \begin{itemize}
        \item Answer relevancy assesses the alignment of the generated response with the user's query, a critical factor for user satisfaction.
        \item \textbf{Dorna with Dorna Embeddings} outperforms other models with a \textbf{score of 0.823}, demonstrating its strength in producing highly relevant responses tailored to Persian-language queries. This indicates the effectiveness of the Dorna model in embedding semantic understanding of user intent in Persian.
        \item The lowest score for this metric is observed with \textbf{GPT-3.5-turbo using Persian-Sentence-Embedding-V3 (0.493)}, suggesting potential challenges in aligning query semantics with response generation when utilizing this embedding.
    \end{itemize}

    \item \textbf{Context Relevancy:}
    \begin{itemize}
        \item Context relevancy measures how well the generated response incorporates broader contextual understanding, ensuring a coherent and comprehensive answer.
        \item \textbf{GPT-3.5-turbo with Persian-Sentence-Embedding-V3} achieves the highest context relevancy score (\textbf{0.223}), demonstrating its relative strength in capturing and incorporating contextual nuances, despite lower performance in other metrics.
        \item \textbf{Dorna with Dorna Embeddings} also performs well (\textbf{0.216}), reflecting its capacity to balance context incorporation with other aspects of performance.
    \end{itemize}
\end{itemize}

\textbf{Observations and Insights:}
\begin{itemize}
    \item \textbf{Trade-offs between Models and Embeddings:}
    \begin{itemize}
        \item \textbf{GPT-3.5-turbo with OpenAI Embeddings} delivers the best performance in faithfulness, reflecting the general applicability and robustness of these embeddings across various datasets.
        \item Conversely, \textbf{Dorna with Dorna Embeddings} exhibits superior performance in answer relevancy, emphasizing the importance of leveraging embeddings specifically designed for Persian text in achieving domain-specific objectives.
    \end{itemize}

    \item \textbf{Performance Variability of Persian-Specific Embeddings:}
    While \textbf{Persian-Sentence-Embedding-V3} demonstrates strengths in context relevancy, its relatively lower performance in faithfulness and answer relevancy indicates the need for further optimization and training on diverse Persian datasets to improve its applicability for information retrieval tasks.
\end{itemize}

\subsection{Quality of Outputs}
To assess the acceptability of the pipeline outputs from a human perspective, we conducted a qualitative evaluation involving 10 reviewers, including members of the University of Isfahan Artificial Intelligence Community and university students. The evaluators rated the generated answers based on clarity, coherence, and overall satisfaction. Feedback indicated that the majority of the answers were clear, well-structured, and effectively addressed the questions, demonstrating high acceptability.

Moreover, the consistency of the answers was noted, with evaluators observing a uniform level of quality across different questions. This consistency highlights the robustness of our pipeline, confirming its ability to produce reliable and high-quality answers suitable for practical use in a university setting.

\section{Conclusion}
\label{sec:conclusion}
In this study, we developed a question-answering pipeline based on Retrieval-Augmented Generation (RAG) using a quantized version of Dorna model, a fine-tuned version of LLaMA-3 on Persian data, and our custom dataset, UniversityQuestionBench (UQB). Our pipeline demonstrated strong performance across three key metrics: faithfulness, answer relevance, and context relevance. The quantitative results were complemented by a qualitative assessment, which confirmed the high acceptability and consistency of the generated answers from a human perspective.

The findings underscore the efficacy of our RAG-based approach in addressing university-level questions, highlighting the potential of fine-tuned language models with context retrieval pipeline and custom datasets in enhancing educational tools.

\section{Future Directions}
\label{sec:future}
Several future contributions are envisioned to further enhance the proposed pipeline's performance and scope, including improvements to the model and dataset. These contributions aim to expand the dataset's diversity, incorporate data from various universities, and establish real-time connectivity with course selection departments for more accurate and up-to-date information. Below are the key future contributions:

\subsection{Contribution 1: Expanding Dataset Diversity}

The first improvement involves expanding the dataset to include more diverse questions and answers, which can increase the robustness of the model across a wider array of academic subjects and contexts. This can be achieved by crowdsourcing data contributions from a broader range of students and institutions, ensuring the dataset contains both valid and new question-answer pairs. By increasing the diversity of the dataset, the model's generalizability and effectiveness in handling various academic queries are expected to improve significantly.

\subsection{Contribution 2: Incorporating Multi-University Data}

To enhance the generalization and versatility of the dataset, it is proposed that question-answer data from multiple universities be incorporated. Each university can have its unique structure, curriculum, and question patterns, which, when combined, create a more generalized dataset suitable for a wide range of academic scenarios. By pooling datasets from multiple institutions, the model will gain exposure to a wider variety of scholarly discourse, which can improve its performance across different contexts and subject areas.

\subsection{Contribution 3: Integrating Real-Time Updates}

Integrating the dataset with course selection departments for real-time updates is crucial to maintaining its relevance and accuracy. This connection ensures that new questions and answers reflecting the most current course offerings and content are continuously added to the dataset. This approach will ensure that the model stays up-to-date with the latest academic developments, enhancing its utility and accuracy for real-time student inquiries.

By implementing these contributions, the pipeline will become more dynamic and capable of handling a wider range of questions in various academic contexts while ensuring up-to-date information is always available to users. These advancements are expected to lead to a more effective and adaptive question-answer system for students and educators alike.

\bibliographystyle{IEEEtran}
\bibliography{IEEEabrv,Ref}

\newpage
\appendix
\label{sec:appendix}
\textbf{Examples of Dataset Questions and Answers:}

Below are sample questions from the dataset and their corresponding answers:

\begin{figure}[h]
    \centering
    \includegraphics[width=1\linewidth]{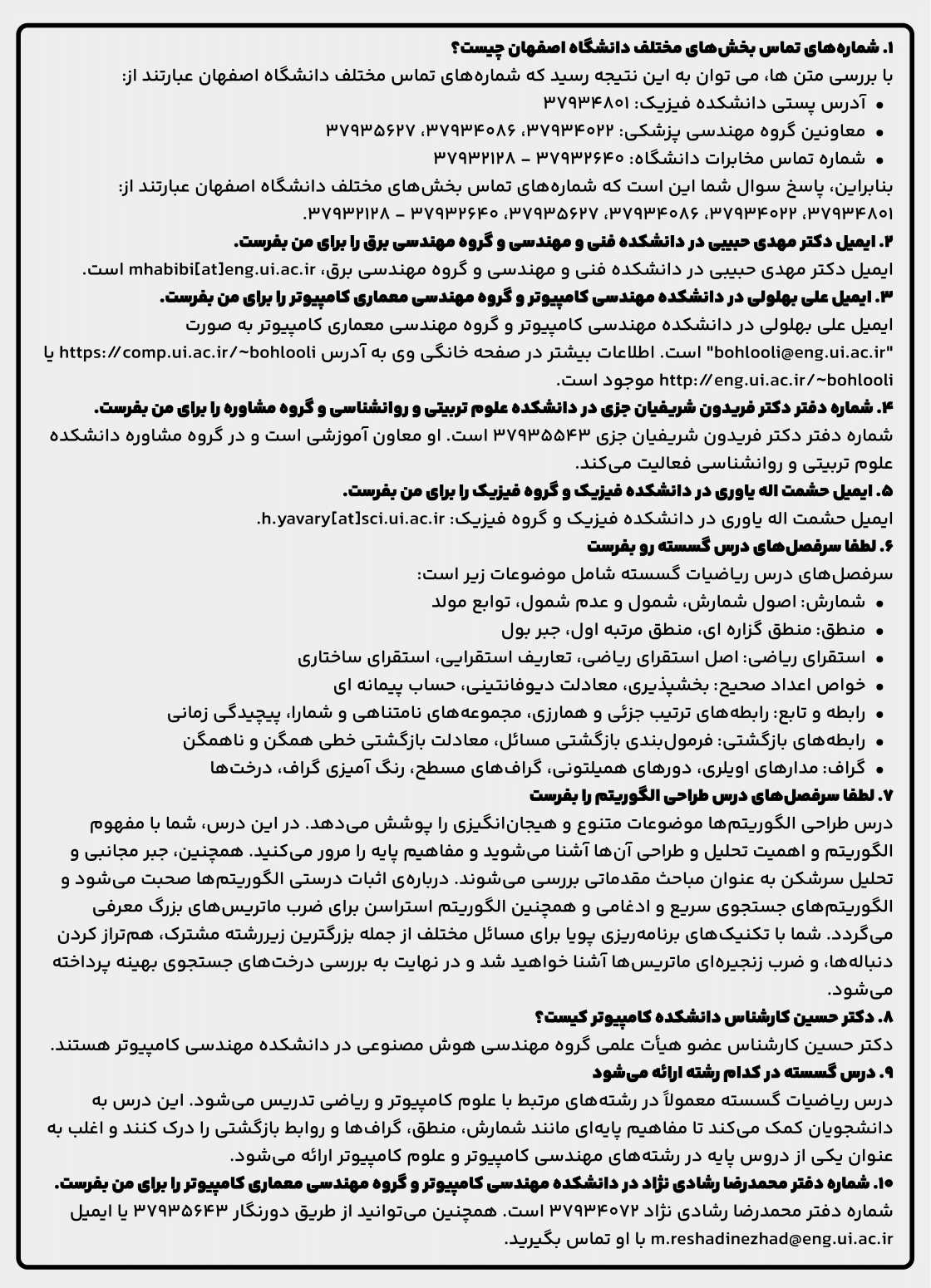}
    \caption{QA samples}
    \label{fig:enter-label}
\end{figure}

\end{document}